\documentclass[%
 twocolumn,
superscriptaddress,
 amsmath,amssymb,
 aps,
]{revtex4-2}
\usepackage{graphicx}% Include figure files
\usepackage{dcolumn}% Align table columns on decimal point
\usepackage{bm}% bold math
\usepackage{xcolor}
\usepackage[caption=false]{subfig}
\usepackage{ulem}

\begin{document}

%\preprint{APS/123-QED}

%\title{Neutron Evaporation Spectra Reveal the $^{60}$Zn Level Density}
\title{Determination of the $^{60}$Zn Level Density from Neutron Evaporation Spectra}

\author{D. Soltesz}
\email{ds149513@ohio.edu}
\affiliation{Institute of Nuclear \& Particle Physics, Department of Physics \& Astronomy, Ohio University, Athens, Ohio 45701, USA}
\author{M.A.A. Mamun}
\affiliation{Institute of Nuclear \& Particle Physics, Department of Physics \& Astronomy, Ohio University, Athens, Ohio 45701, USA}
\author{A.V. Voinov}
\affiliation{Institute of Nuclear \& Particle Physics, Department of Physics \& Astronomy, Ohio University, Athens, Ohio 45701, USA}
\author{Z. Meisel}
\email{meisel@ohio.edu}
\affiliation{Institute of Nuclear \& Particle Physics, Department of Physics \& Astronomy, Ohio University, Athens, Ohio 45701, USA}
\author{B.A. Brown}
\affiliation{Department of Physics \& Astronomy and National Superconducting Cyclotron Laboratory, East Lansing, Michigan 48824, USA}
\author{C.R. Brune}
\affiliation{Institute of Nuclear \& Particle Physics, Department of Physics \& Astronomy, Ohio University, Athens, Ohio 45701, USA}
\author{S.M. Grimes}
\affiliation{Institute of Nuclear \& Particle Physics, Department of Physics \& Astronomy, Ohio University, Athens, Ohio 45701, USA}
\author{H. Hadizadeh}
\affiliation{Institute of Nuclear \& Particle Physics, Department of Physics \& Astronomy, Ohio University, Athens, Ohio 45701, USA}
\author{M.Hornish}
\affiliation{Institute of Nuclear \& Particle Physics, Department of Physics \& Astronomy, Ohio University, Athens, Ohio 45701, USA}
\author{T.N. Massey}
\affiliation{Institute of Nuclear \& Particle Physics, Department of Physics \& Astronomy, Ohio University, Athens, Ohio 45701, USA}
\author{J.E. O'Donnell}
\altaffiliation{Deceased.}
\affiliation{Institute of Nuclear \& Particle Physics, Department of Physics \& Astronomy, Ohio University, Athens, Ohio 45701, USA}
\author{W.E. Ormand}
\affiliation{Lawrence Livermore National Laboratory, P.O. Box 808, L-235, Livermore, California 94551, USA}

\begin{abstract}
Nuclear reactions of interest for astrophysics and applications often rely on statistical model calculations for nuclear reaction rates, particularly for nuclei far from $\beta$-stability. However, statistical model parameters are often poorly constrained, where experimental constraints are particularly sparse for exotic nuclides. For example, our understanding of the breakout from the NiCu cycle in the astrophysical $rp$-process is currently limited by uncertainties in the statistical properties of the proton-rich nucleus $^{60}$Zn. We have determined the nuclear level density of $^{60}$Zn using neutron evaporation spectra from $^{58}{\rm Ni}(^3{\rm He},n)$  measured at the Edwards Accelerator Laboratory. We compare our results to a number of theoretical predictions, including phenomenological, microscopic, and shell model based approaches. Notably, we find the $^{60}{\rm Zn}$ level density is somewhat lower than expected for excitation energies populated in the $^{59}{\rm Cu}(p,\gamma)^{60}{\rm Zn}$ reaction under $rp$-process conditions. This includes a level density plateau from roughly 5--6~MeV excitation energy, which is counter to the usual expectation of exponential growth and all theoretical predictions that we explore. A determination of the spin-distribution at the relevant excitation energies in $^{60}{\rm Zn}$ is needed to confirm that the Hauser-Feshbach formalism is appropriate for the $^{59}{\rm Cu}(p,\gamma)^{60}{\rm Zn}$ reaction rate at X-ray burst temperatures.
\end{abstract}

\maketitle

\section{\label{sec:level1} Introduction}
Atomic nuclei have long been known to be well described by statistical properties at excitation energies with a relatively dense spacing of nuclear levels~\cite{BohrScience1937,Weisskopf1937}. The number of levels per excitation energy $E^{*}$, referred to as the nuclear level density $\rho$, is of particular importance. For instance, $\rho$ determines the number of states accessible in the decay of a compound nucleus via a particular particle or $\gamma$-emission channel. While $\rho$ has been well characterized for nuclei along the valley of $\beta$-stability~\cite{RIPL_IAEA_handbook}, experimental challenges have made constraints for short-lived nuclides virtually absent. This is troubling considering the key role statistical estimates of nuclear reaction rates play in calculations of astrophysical phenomena and for nuclear applications~\cite{Arcones,Empire}.
 
  The level density enters into the calculation of the nuclear reaction cross section through the  Hauser-Feshbach (HF) formalism~\cite{Wolfenstein,HF} via $\sigma_{\rm HF}\propto\lambda^{2}(\mathcal{T}_{\rm form}\mathcal{T}_{{\rm decay,}i})/\Sigma_{j }\mathcal{T}_{{\rm decay,}j}$, where $\lambda$ is the de Broglie wavelength for the entrance channel and $\mathcal{T}$ are the transmission coefficients that describe the probability for a particle or photon, which defines the channel, being emitted from (``decay") or absorbed by (``form") a nucleus. The $\mathcal{T}$ for decay channels in the preceding equation, $\mathcal{T}_{{\rm decay},i}$ and all other open decay channels included in the sum $\Sigma_{j}\mathcal{T}_{{\rm decay},j}$, are actually a sum over $\mathcal{T}$ to individual discrete states added to an integral over $\mathcal{T}$ to levels in a higher-excitation energy region described by $\rho$.

A variety of methods can be used to obtain $\rho$. Primary techniques include a direct determination of level spacings by level counting or neutron resonance spacings~\cite{Raman83,vonEgidy}; proton scattering at extreme forward angles~\cite{Polt2014}; Ericson fluctuations~\cite{Eric60}; $\beta$-delayed particle spectrum fluctuations~\cite{Giov00}; the regular, inverse kinematics, or $\beta$ Oslo method
%, although this method requires an estimate for the neutron resonance spacing off of stability
~\cite{Schiller,Ingeberg,Spyrou,Liddick}; and particle evaporation spectra ~\cite{Byun,Ramirez13,Voinov}. Of the techniques that can be used on exotic nuclides, neutron resonance spacings and $\beta$-Oslo measurements require estimates of the spin distribution of excited states in order to obtain a total level density. Furthermore, the latter technique requires normalization to neutron resonance spacings, which generally must be calculated for exotic nuclides. Meanwhile, particle evaporation spectra provide a nearly model-independent level density extraction, albeit presently limited to nuclides a few nucleons from stability.
 
To date, constraints off stability (which do not rely on theoretical normalizations) are extremely limited. Results for $\rho$ from counting discrete states show hints of a reduction in $\rho$ for increasingly exotic nuclides~\cite{AlQur01,AlQur}. The purpose of this work is to provide a constraint for $\rho$ of a nucleus several nucleons from stability, using a model-independent measurement of $\rho$.

Here, we focus on $\rho$ for $^{60}\rm{Zn}$ due to its role in the astrophysical rapid proton capture ($rp$-) process. The $rp$-process powers type-I X-ray bursts, thermonuclear explosions that occur on the surfaces of accreting neutron stars~\cite{Meis18}. $^{60}{\rm Zn}$ plays a key role in the NiCu cycle of the $rp$-process, where transmutation of $^{59}{\rm Cu}$ via a $(p,\alpha)$ reaction stalls nuclear energy generation, while a $(p,\gamma)$ reaction, which depends on $\rho$ of $^{60}{\rm Zn}$, enables the $rp$-process to proceed~\cite{VanWormer}. X-ray burst model calculations have shown that the shape of the burst light curve is particularly sensitive to the $^{59}{\rm Cu}(p,\gamma)^{60}{\rm Zn}$ reaction rate, which may impact constraints on the properties of ultradense matter~\cite{Cyburt16,Meis19}.

We report the first measurement of $\rho$ for $^{60}$Zn, obtained using neutron evaporation spectra from $^{58}{\rm Ni}(^3{\rm He},n)^{60}{\rm Zn}$ measured at the Edwards Accelerator Laboratory at Ohio University~\cite{Meis17}. We describe our experiment in Section~\ref{sec::level2} and analysis in Section~\ref{sec::level3}. In Section~\ref{sec::level4} we present our results, followed by a comparison to theoretical level density models and discussion of the implications for astrophysics in Section~\ref{sec::level5}. We conclude in Section~\ref{sec::level6}, including some comments on next steps for constraining the $^{59}{\rm Cu}(p,\gamma)$ reaction rate for the $rp$-process.

\section{\label{sec::level2} Experiment Set-up}
Measurements were performed at the Edwards Accelerator Laboratory using the beam swinger and neutron time-of-flight (TOF) tunnel~\cite{Meis17,Finlay}. A beam of $^{3}{\rm He}$ was chopped and bunched into $\sim$3~ns bunches with a frequency of 2.5~MHz, accelerated to 10~MeV using the 4.5~MV T-type tandem Pelletron, and impinged on an 0.525$\pm$.053~mg/cm$^{2}$ $^{58}{\rm Ni}$ target with an average intensity of $\sim20$~nA. Due to energy loss within the $^{58}$Ni target, the average energy of the $^3$He beam at the center of the target was 9.93~MeV. Incident beam current was collected by integrating charge deposited in the target chamber. Neutrons were detected at an angle of 105$^{\circ}$ by three 5.08~cm thick, 12.7~cm diameter NE213 liquid organic scintillators, located at a distance of $L_{\rm path}=5$~m from the target within the neutron TOF tunnel. The TOF start was provided by a detection within an NE213 detector, with TOF stop provided by charge capacitively deposited in a beam pick-off located upstream of the target chamber.

To identify contributions from beam-induced background to the $^{58}{\rm Ni}(^{3}{\rm He},n)$ spectrum, measurements were performed using a thin carbon target and an empty target frame. Additionally, measurements with all targets were taken at an angle of 115$^{\circ}$ to exploit the kinematic shifts of neutron peaks at discrete energies.

\begin{figure}
    \begin{center}
    \includegraphics[width=\columnwidth]{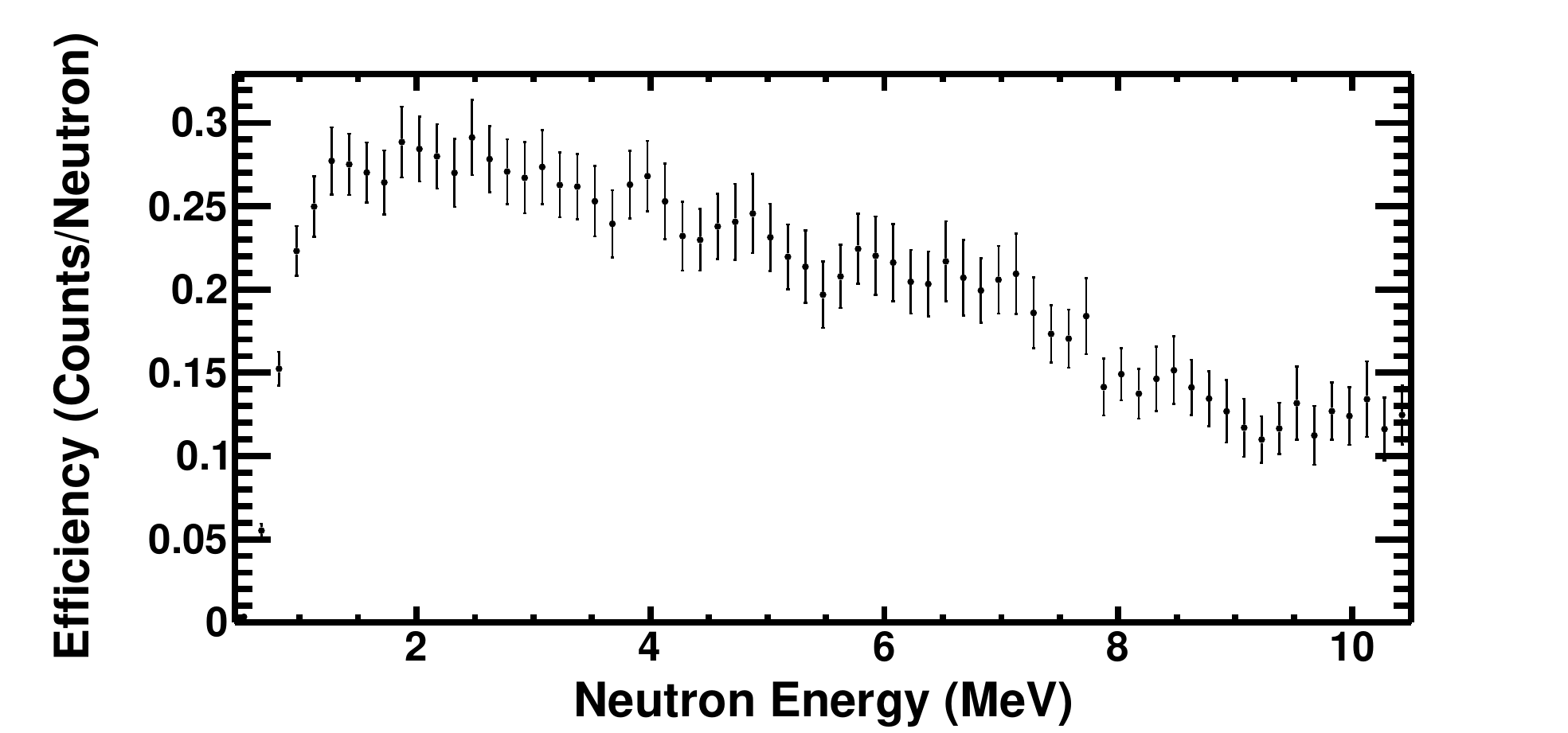}
    \caption{Intrinsic NE213 neutron detection efficiency determined using $^{27}{\rm Al}(d,n)$.}
    \label{Eff}
    \end{center}
\end{figure}

Detector efficiency calibrations were performed by measuring a standard. Here, neutrons from $^{27}{\rm Al}(d,n)$ were measured at 120$^{\circ}$ for 7.44~MeV deuterons impinging on a thick Al target~\cite{AlDn}. Our neutron detection efficiency is shown in Figure~\ref{Eff}.%The spectrum from this reaction and philosophy behind this approach for the neutron detection efficiency are presented in Ref.~\cite{AlDn}.

\section{\label{sec::level3} Analysis}

\subsection{Neutron Differential Cross Section Determination}
\label{ssec:DCSdetermination}

Neutrons were identified using TOF and pulse shape discrimination, providing $n$-$\gamma$ separation down to the NE213 threshold near 1~MeV. A linear time calibration was performed using each spectrum's $\gamma$-ray peak and the neutron peak associated with populating the ground state of $^{14}{\rm O}$ via $^{12}{\rm C}(^{3}{\rm He},n)$. The neutron laboratory frame energy is $E_{n}=m_{n}\left(\sqrt{1/(1-\beta_{n}^{2})}-1\right)$, where $m_{n}$ is the neutron mass, $\beta_{n}=(L_{\rm path}/{\rm TOF})/c$, and $c$ is the speed of light.

The constant time-independent background from random coincidences triggered by background radiation was determined using the superluminal region of each spectrum. This average background level was determined individually for each NE213 detector and subtracted from the corresponding neutron spectrum. Neutrons from reactions on target contaminants were removed from the differential cross section as described below. 

The approach of Ref.~\cite{AlDn} was used to determine the neutron detection efficiency, carrying a statistical uncertainty along with a 5$\%$ systematic uncertainty. By comparing our measured $^{27}{\rm Al}(d,n)$ spectrum to the standard spectrum determined in that work, we determine the product of the geometric and intrinsic neutron detection efficiency. 

\begin{figure}
    \begin{center}
    \includegraphics[width=\columnwidth]{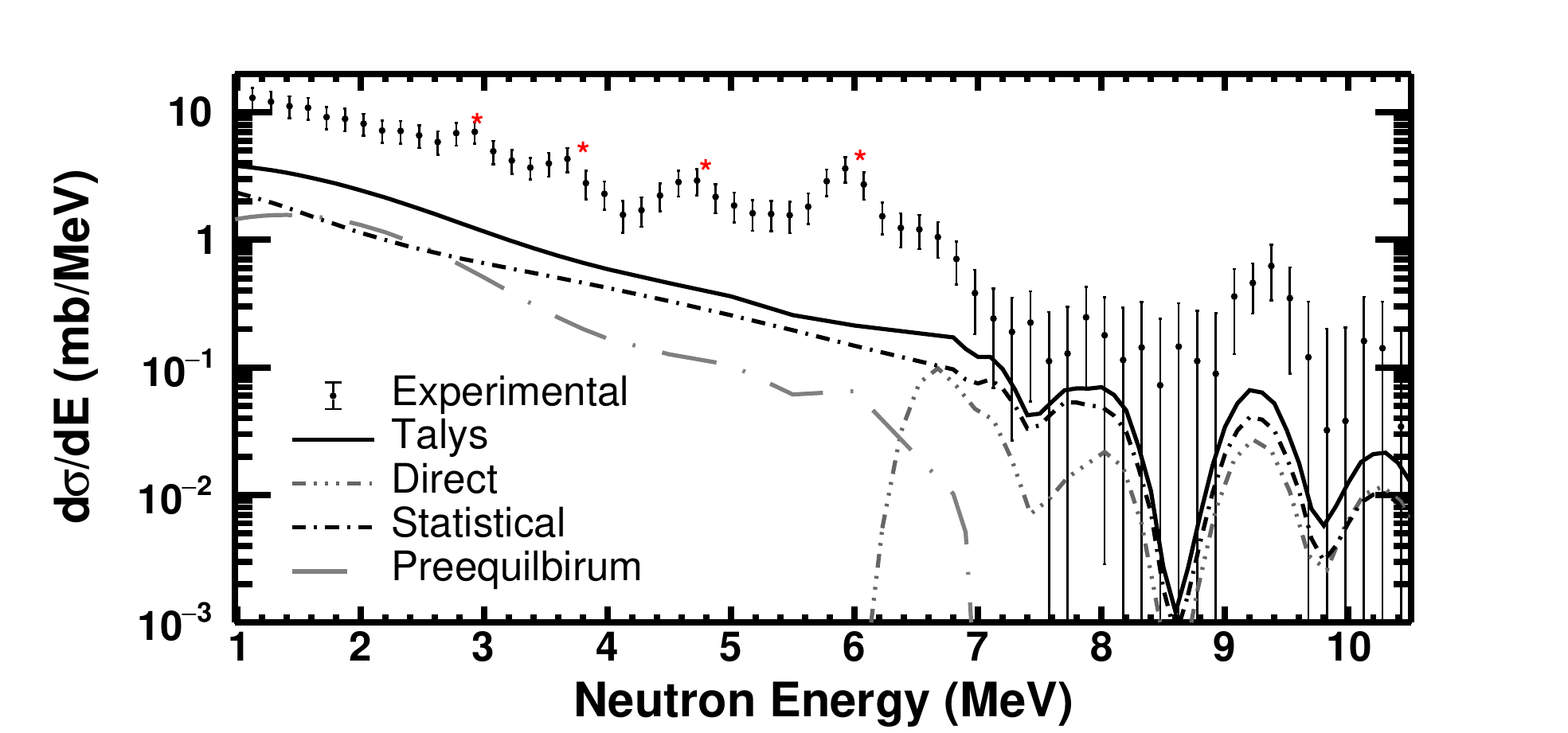}
    \caption{Measured (points) $d\sigma/dE$  for $^{58}{\rm Ni}(^{3}{\rm He},n)$ at 105$^{\circ}$ 
    compared to default {\tt TalysV1.8} calculations, including a breakdown of the total $d\sigma/dE$ by reaction mechanism. Asterisks indicate contamination from $(^{3}{\rm He},n)$ on carbon and oxygen.}
    \label{ContamCross}
    \end{center}
\end{figure}

Following the time calibration and efficiency correction, the spectra of the three NE213 detectors were combined. The differential cross section $d\sigma/dEd\Omega=N_{\rm det}/\left(N_{\rm beam}\mathfrak{n}_{A}\epsilon_{\rm tot}\right)$, where $N_{\rm det}$ is the number of detected neutrons within an energy interval around energy $E$, $N_{\rm beam}$ is the number of beam particles incident on the target, $\mathfrak{n}_{A}$ is the areal atomic density of the target, and $\epsilon_{\rm tot}$ is the total (geometric$\times$intrinsic) neutron detection efficiency at energy $E$. $N_{\rm beam}$ was measured using charge integration on the target, carrying a 3\% uncertainty. Figure~\ref{ContamCross} compares our $d\sigma/dE$, where we have multiplied the measured $d\sigma/dEd\Omega$ by 4$\pi$~sr, in the center-of-mass frame to calculations performed with the code {\tt Talys} version 1.8 ({\tt TalysV1.8})~\cite{talys} using the default model parameters.

\begin{figure*}
    \centering
    \subfloat{\includegraphics[width=\columnwidth]{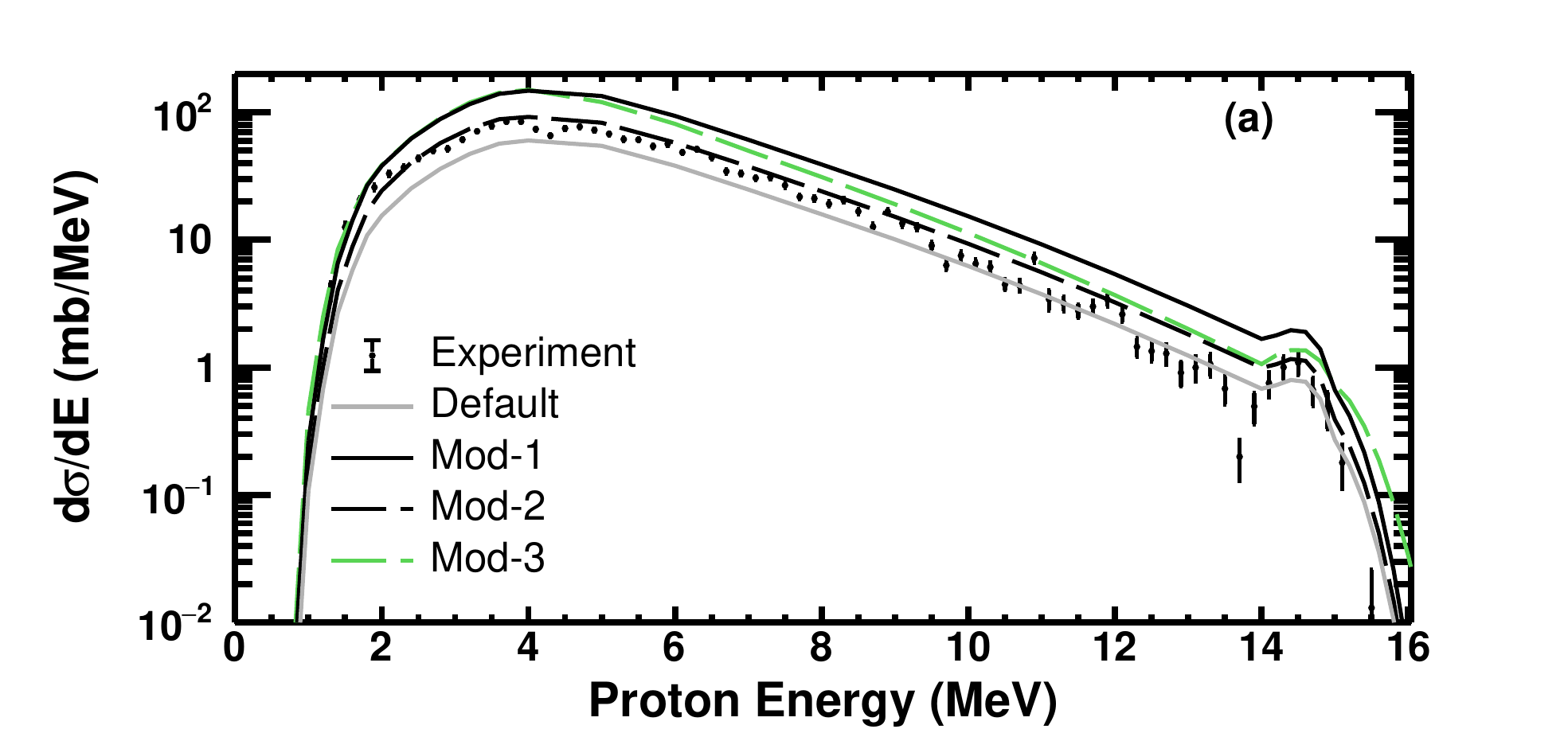}}
    \subfloat{\includegraphics[width=\columnwidth]{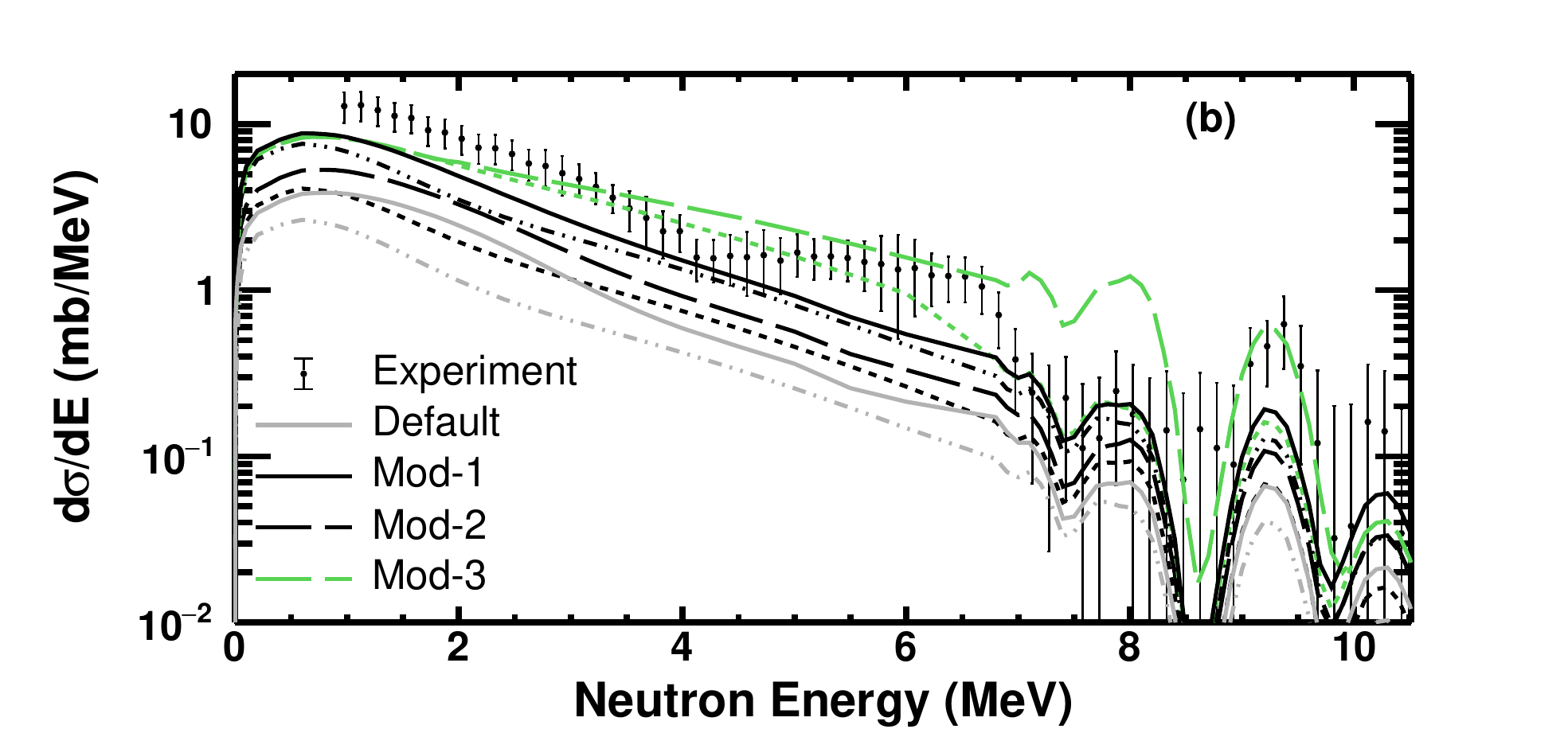}}
    \caption{$^{3}{\rm He}+^{58}{\rm Ni}$ proton (a) and neutron (b) spectra compared to results from model calculations we describe as Mod-1, Mod-2, and Mod-3 in the text. Solid and long-dashed lines show the total $d\sigma/dE$, while the short-dashed and dot-dash lines show the contribution from the statistical reaction mechanism to the long-dashed and solid lines,respectively}.
    \label{OptMods}
\end{figure*}

The $^{58}{\rm Ni}(^{3}{\rm He},n)$ $d\sigma/dE$ is expected to have relatively smooth behavior up to $E_{\rm n}\sim7$~MeV, at which point excited states in $^{60}{\rm Zn}$ become sparse enough that the neutron spectrum will be characterized by discrete peaks. Our data largely follow this trend; however, several prominent discrete peaks are apparent in the continuum region. We do not attribute these to reactions on $^{58}{\rm Ni}$ favoring particular states in a direct reaction mechanism, as this should be disfavored by the relatively low reaction energy and backward angles for neutron detection. Instead, these peaks in the continuum region can be understood as background from $^{12}{\rm C}(^{3}{\rm He},n)$ and $^{16}{\rm O}(^{3}{\rm He},n)$ reactions on contamination in the target and collimator upstream of the target. We confirmed this identification using our measurements on the carbon target and empty target frame, including checking the kinematic shift of the peaks for the 115$^{\circ}$ measurement angle relative to the 105$^{\circ}$ angle, along with calculations using {\tt TalysV1.8}. Oxygen and carbon background peaks were removed by fitting each peak in the continuum with a Gaussian summed with a polynomial. The Gaussian contributions were then subtracted from the measured $d\sigma/dE$.

\subsection{Level Density Determination}
\label{ssec:LDextraction}

The general approach for our level density determination is briefly summarized here and elaborated upon in the subsequent subsections. Calculations are performed for the $^{58}{\rm Ni}(^{3}{\rm He},n)$ $d\sigma/dE$ assuming some theoretical $\rho$ for $^{60}{\rm Zn}$. The ratio between the measured and calculated $d\sigma/dE$ is determined and then multiplied by $\rho$ used in the calculation, taking advantage of the proportionality $d\sigma(E)/dE\propto\rho(E^{*})$, where $E^{*}$ is the excitation energy populated by a neutron of energy $E$~\cite{Blat52}. The absolute $\rho$ is determined by performing a normalization to match the theoretical $d\sigma/dE$ populating the low-lying excitation energy region of $^{60}{\rm Zn}$, where all levels are thought to be known. %This process is then iterated, beginning each step with the previously extracted $\rho$ as the theoretical input for the calculation of $d\sigma/dE$, until the extracted $\rho$ converges.

The detailed description of this process is as follows. 

\subsubsection{$d\sigma/dE$ Calculation}

The level density extraction begins with a theoretical estimate for $d\sigma/dE$.
While the final extracted $\rho$, $\rho_{\rm expt}$, is relatively insensitive to the initial theoretical $\rho$, $\rho_{\rm th}$, it is not as clear that our result will be as insensitive to other assumptions in the model calculation of $d\sigma/dE$. In particular, there is considerable uncertainty in the optical model potential (OMP) for $^{3}{\rm He}$ and the pre-equilibrium contribution to $d\sigma/dE$, where models in general struggle to reproduce $^{3}{\rm He}$-induced reaction data~\cite{Kalbach2005,Duis2016}. The $^{3}{\rm He}$ OMP will impact the overall magnitude of the $(^{3}{\rm He},n)$ cross section, but this impact is removed in the $\rho$ renormalization (described below). Here, the concern with the $^{3}{\rm He}$ OMP is the impact on the populated spin distribution for the compound nucleus, $^{61}{\rm Zn}$, which will modify the distribution of centrifugal barriers for outgoing neutrons, and thereby influence the $(^{3}{\rm He},n)$ spectrum. Similarly, the pre-equilibrium contribution can modify the neutron spectrum and any suspected contributions from this reaction mechanism must be subtracted from the data in order to maintain the validity of the proportionality $d\sigma(E)/dE\propto\rho(E^{*})$.

The $^{3}{\rm He}$ OMP and pre-equilibrium model parameters were selected by reproducing measured $^{58}{\rm Ni}(^{3}{\rm He},p)$ data with {\tt TalysV1.8}, while checking the influence on the associated $^{58}{\rm Ni}(^{3}{\rm He},n)$ spectrum (see Figure~\ref{OptMods}). The $(^{3}{\rm He},p)$ data, which are the focus of a separate work, are from a measurement following the approach of Ref.~\cite{VoinovHF}. Briefly, these $(^{3}{\rm He},p)$ spectra were produced using a 10~MeV beam of $^{3}{\rm He}$ by employing charged-particle time-of-flight and energy loss with a silicon surface barrier detector located at a flight path of 1~m and angle of 112$^{\circ}$. We note that these data are consistent with the lower statistics measurement of Ref.~\cite{Agui13}. The benefit of including the proton spectrum in the present analysis is that these data are sensitive to the formation cross section of the $^{61}{\rm Zn}$ compound nucleus, but not the neutron-decay daughter $^{60}{\rm Zn}$. Since the $(^{3}{\rm He},p)$ channel dominates over the $(^{3}{\rm He},n)$ channel at these energies, reproducing the magnitude of the former is a strong indication that we are reproducing the $^{3}{\rm He}+^{58}{\rm Ni}$ fusion cross section.

We explored a variety of modifications to the $^{3}{\rm He}$ OMP and pre-equilbrium models, focusing in Figure~\ref{OptMods} on the subset that best reproduced the shape of the proton and neutron spectra, while using the constant temperature level density model (see Section~\ref{sec::level4}) and the mean beam energy in the center of the target. We refer to these modifications as Mod-1,  Mod-2, and Mod-3, and compare to the results from using the default {\tt TalysV1.8} inputs for context. The default $^{3}{\rm He}$ OMP is essentially a weighted combination~\footnote{The real and imaginary central and spin-orbit potential depths, as well as the potential radius and diffuseness for each potential component, are weighted by proton number $Z$ and neutron number $N$ for $^{3}{\rm He}$.} of the proton and neutron OMPs from Ref.~\cite{Koni03}, where in this case local OMPs are available. The default pre-equilibrium model is a two-component exciton model~\cite{KONING2004} supplemented by contributions from nucleon transfer and knock-out~\cite{Kalbach2005}.
For Mod-1, we use the 
%built-in empirical normalization of
%the transmission coefficients calculated for the $^{3}{\rm He}$ OMP
$\mathcal{T}_{^{3}{\rm He}}$ based on the systematics established by Ref.~\cite{Trip96,Trip97}. For Mod-2, we increase the leading coefficient of the surface absorption term for the $^{3}{\rm He}$ OMP by a factor of 10. For Mod-3, we use the $\mathcal{T}_{^{3}{\rm He}}$ from systematics referred to in Mod-1 and a multi-step direct and compound pre-equilibrium model~\cite{Fesh80} decreased by the scale factor 0.3. We find that Mod-2 most satisfactorily describes the $(^{3}{\rm He},p)$ data. The increased absorption implied by Mod-2 is consistent with the $^{3}{\rm He}$\,+\,$^{58}{\rm Ni}$ fusion cross section enhancement found by Ref.~\cite{Agui13}, where near 10~MeV Mod-2 results in $\sigma_{\rm fus.}=250$~mb, as compared to 220~mb in Ref.~\cite{Agui13}. Therefore, we use the Mod-2 $^{3}{\rm He}$ OMP and pre-equilibrium combination for the remainder of the analysis. The corresponding impact on the $(^{3}{\rm He},n)$ spectrum is shown in Figure~\ref{OptMods}b.

\subsubsection{$\rho(E^{*})$ Extraction}

Determining $\rho(E^{*})$ from a particle evaporation spectrum is based on the concept that $d\sigma(E)/dE$ will depend on the number of states near excitation energy $E^{*}$ that will be populated by a particle evaporating from the compound nucleus with an energy near $E$~\cite{Vonach83,Wallner95,VoinovGe}.

The proportionality $d\sigma(E)/dE\propto\rho(E^{*})$ is only valid for reaction products created via the statistical reaction mechanism and relies on accurate estimates for $\mathcal{T}$. The analysis presented in the prior subsection served to determine the non-statistical contributions to our $(^{3}{\rm He},n)$ spectrum, as well as $\mathcal{T}_{^{3}{\rm He}}$. $\mathcal{T}_{n}$ used here result from the neutron OMP of Ref.~\cite{Koni03}, though we found little difference when employing the JLM potentials of Refs.~\cite{Baug01,Gori07}.

\begin{figure}
   \centering
    \includegraphics[width=\columnwidth]{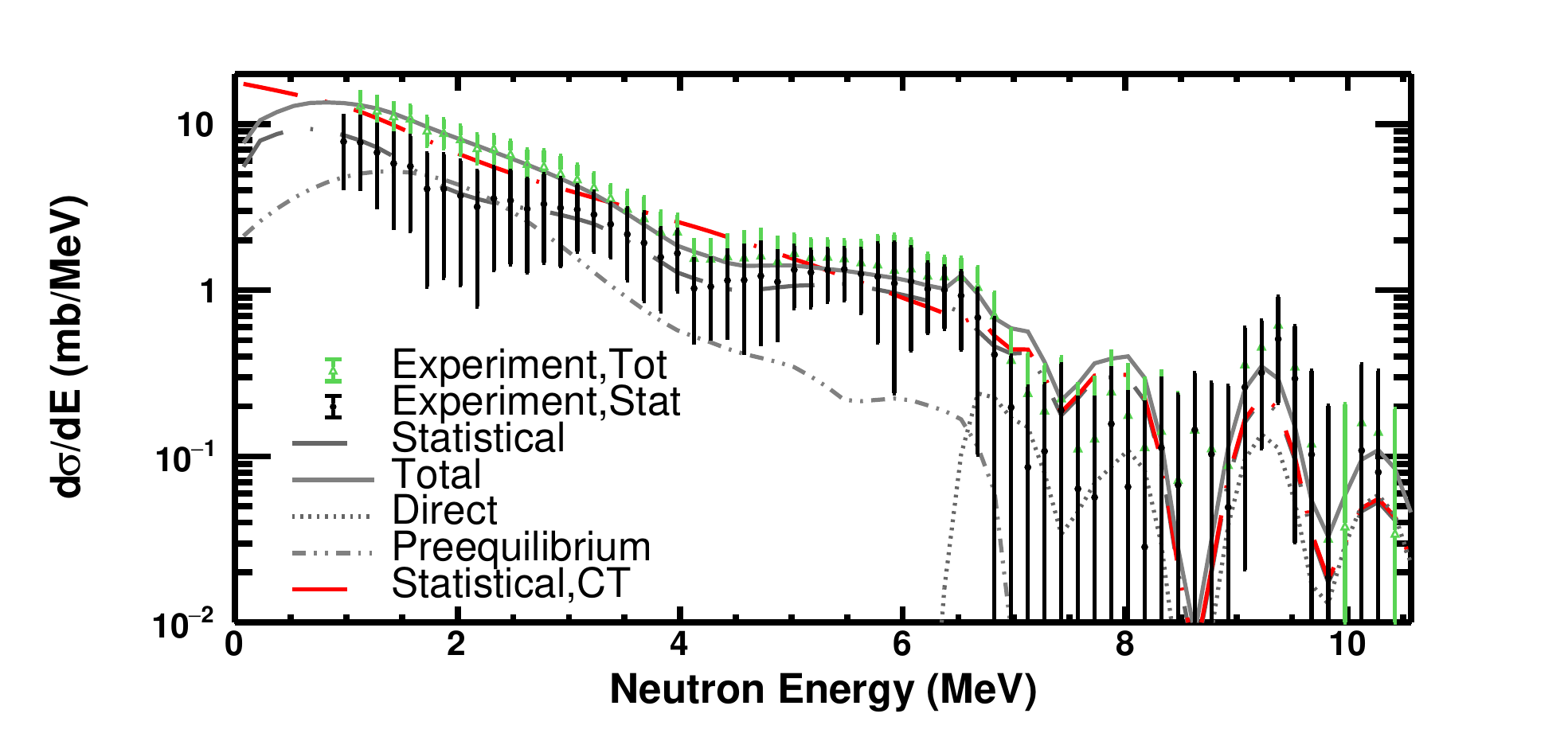}
    \caption{Measured $^{58}{\rm Ni}(^{3}{\rm He},n)$ differential cross section before (open triangles) and after (filled triangles) subtracting calculation results for non-statistical contributions. Calculation results for the total differential cross section using $\rho_{\rm expt}$ (solid gray line) are shown, along with the contributions to the total from the direct (dotted line), pre-equilibrium (dot-dot-dash line), and statistical (dot-dash line) components. For comparison, the statistical component of the differential cross section calculated using $\rho_{\rm CT}$ is also shown.}
    \label{FinalCS}
\end{figure}

Essentially, the experimental level density $\rho_{\rm expt}$ is determined by scaling a theoretical level density $\rho_{\rm th}$ used to calculate $d\sigma/dE|_{\rm th}$ by the ratio between the experimental and theoretical $d\sigma/dE$:
\begin{equation}
   \rho_{\rm expt}(E^{*})\propto\rho_{\rm th}(E^{*})\frac{d\sigma(E)/dE|_{\rm expt}}{d\sigma(E)/dE|_{\rm th}},
    \label{eqn:RhoEqn}
\end{equation}
where $E^{*}$ and $E$ are related by kinematics and here the $d\sigma/dE$ refer only to the statistical component~\cite{Vonach83}. %To isolate the statistical component of $d\sigma/dE|_{\rm expt}$, we subtract the calculated direct and pre-equilibrium contributions to $d\sigma/dE$. Prior to subtraction, these contributions are first scaled by the same factor required for the total calculated $d\sigma/dE$ to maximize agreement with the total measured $d\sigma/dE$. %This is performed iteratively, as described below.
The detailed $\rho$ extraction steps are as follows:

\begin{enumerate}
\item As an initial estimate, the constant temperature (CT) phenomenological level density~\cite{Gilb65} $\rho_{\rm CT}$ is used for $\rho_{\rm th}$, where
\begin{equation}
\rho_{\rm CT}(E^{*})=\frac{1}{T}\exp\left(\frac{E^{*}-\delta E}{T}\right),
\label{eqn:CT}
\end{equation}
and the nuclear temperature $T$ and energy shift $\delta E$ are from global parameter systematics of Ref.~\cite{RIPL_IAEA_handbook}. Using this $\rho_{\rm th}$, $d\sigma/dE|_{\rm th}$ (the statistical contribution to the theoretical $d\sigma/dE$) is calculated. {\tt TalysV1.8} results are reported separately for the continuum and discrete level region, where the latter includes by default the ten lowest-lying excited states. We apply our experimentally determined neutron energy resolution to $d\sigma/dE$ calculated for the discrete-level region in order to compare to $d\sigma/dE|_{\rm expt}$. The neutron energy resolution was calculated as
\begin{equation}
\Delta{E_{n}} = \sqrt{\Delta{E}^2_{\rm loss}+\Delta{E}^2_{\rm TOF}},
\end{equation}
where $\Delta{E}_{\rm loss}$ is the energy loss within the target calculated using LISE++~\cite{LISE++}. We ignore energy resolution contributions due to energy straggling in the target, as  $\Delta{E}_{\rm strag}<<\Delta{E}_{\rm loss}$. $\Delta{ E}_{\rm TOF}$ is the uncertainty in the neutron energy from the TOF:
\begin{equation}
    \Delta{E}_{\rm TOF} = 2{E_n}\sqrt{\left(\frac{L_{\rm path}}{D}\right)^2+\left(\frac{\Delta{t}}{\rm TOF}\right)^2},
\end{equation}
where $D$ and $\Delta{t}$ are the detector thickness and the bunch length of the beam packets, respectively. For instance, at $E_{n}=7.050$~MeV, $\Delta E_{\rm TOF}=0.293$~MeV, and $\Delta E_{n}=0.323$~MeV. %\sout{Blending the continuum results to the discrete level region results creates a discontinuity around $E=6-6.75$~MeV; therefore, we do not extract $\rho_{\rm expt}$ for $E^{*}$ corresponding to those $E$}}.

\item The direct and pre-equilibrium contributions to the calculated $d\sigma/dE$, scaled along with the statistical contribution to minimize $\chi^{2}$ between the total calculated $d\sigma/dE$ and the measured $d\sigma/dE$, are subtracted from the measured total $d\sigma/dE$ to arrive at $d\sigma/dE|_{\rm expt}$ (which is the statistical contribution to the measured $d\sigma/dE$). We use $\Delta\chi^{2}=\chi^{2}-\chi^{2}_{\rm min}$ to determine the uncertainty~\cite{Pres92} in the scale factor and propagate this through as a contribution to the systematic uncertainty in $\rho_{\rm expt}$. In this analysis, $\chi^2=8.03$, where $\Delta\chi^2 = 1.00$ for a 68$\%$ confidence interval.

\item We normalize $d\sigma/dE|_{\rm th}$ to match the integral of $d\sigma/dE|_{\rm expt}$ over the region corresponding to $E^{*}=0-4$~MeV, which is where all discrete levels in $^{60}{\rm Zn}$, and therefore the absolute $\rho$, are thought to be known~\cite{RIPL_IAEA_handbook}. This normalization is sensitive to uncertainties in $d\sigma/dE|_{\rm expt}$ where the discrete level region is populated. As such, to account for this normalization uncertainty, we repeat this step 1,000 times, each time perturbing the $d\sigma/dE|_{\rm expt}$ in this region randomly within the one standard deviation uncertainty of each data point. The normalization factor $\mathcal{F}=0.795$ and its uncertainty $\delta F=0.149$ are the mean and standard deviation, respectively, of the normalization factor distribution.

\item The level density is determined by
\begin{equation}
  \rho_{\rm expt}(E^{*})=\rho_{\rm th}(E^{*})\frac{d\sigma(E)/dE|_{\rm expt}}{d\sigma(E)/dE|_{\rm th}}\mathcal{F} 
 \label{eqn:RhoFinal}
\end{equation}

\item The level density from the previous step, $\rho_{\rm expt}$, is now used as the the theoretical level density $\rho_{\rm th}$ in step 1, in lieu of $\rho_{\rm CT}$. The levels are distributed by spin $J$, $\rho(E^{*},J)=P(J)\rho(E^{*})$, assuming the Bethe spin-distribution~\cite{Beth36}, which agrees reasonably well with measurements and more sophisticated calculations~\cite{Egidy,Grim16,Orma19}, where the probability of a given $J$ is 
\begin{equation}
    P(J)=\frac{2J+1}{2\sigma^{2}}\exp\left(\frac{-(J+1/2)^{2}}{2\sigma^{2}}\right)
\end{equation}
and $\sigma^{2}$ is the spin-cutoff parameter. 

The $E^{*}$-dependent $\sigma^{2}$ is determined in {\tt Talysv1.8} from a linear interpolation between $\sigma^{2}$ calculated using the known $J$ for levels with $E^{*}\leq4$~MeV, $\sigma_{{\rm disc}}^{2}$, and the Fermi gas estimate $\sigma_{\rm FG}^{2}$ evaluated at the neutron separation energy, $S_{n}$. For $E^{*}\leq4$~MeV, $\sigma_{{\rm disc}}^{2}$ is represented by
\begin{equation}
    \sigma_{\rm disc}^{2}=\frac{\sum{ J_{i}(J_{i}+1)(2J_{i}+1)}}{3\sum{2J_{i}+1}},
    \label{eqn:SigDisc}
\end{equation}
where the sum runs over all levels in the discrete level region, and
\begin{equation}
    \sigma_{\rm FG}^{2}\sim0.04A^{7/6}\sqrt{E^{*}},
    \label{eqn:SigFG}
\end{equation}
where $A$ is the mass number. The exact form is available in the {\tt Talys} manual.

Ref.~\cite{Grim16} found roughly a factor of two variability amongst various predictions and experimental constraints for $\sigma^{2}$ in our $E^{*}$ and $A$ region (see also Figure~\ref{fig:SpinDist}). Therefore, we also investigated performing the $\rho$ extraction described in this subsection for a factor of two increase and factor of two decrease in $\sigma^{2}$. Additionally, we explored using the empirical $\sigma^{2}$ determined by Ref.~\cite{vonEgidy},
where $\sigma^{2}\propto (E^{*})^{0.312}$. The results from using alternative $\sigma^{2}$ parameterizations were used to assess the uncertainty contribution to $\rho_{\rm expt}$ from our choice of $\sigma^{2}$.

\item We calculate $d\sigma/dE$ using the newly extracted $\rho_{\rm expt}$ as the input $\rho$ and compared to the data (in Figure~\ref{FinalCS}) to confirm that using the extracted $\rho$ describes the data.
%The newly extracted $\rho_{\rm expt}$ from repeating steps 1-5 is compared to the $\rho$ used as $\rho_{\rm th}$ for the current iteration. If the two have not converged, a new iteration is started. For this analysis, three iterations was sufficient.

\end{enumerate}
   
\section{\label{sec::level4} Results}

\begin{figure}
    \centering
    \includegraphics[width=\columnwidth]{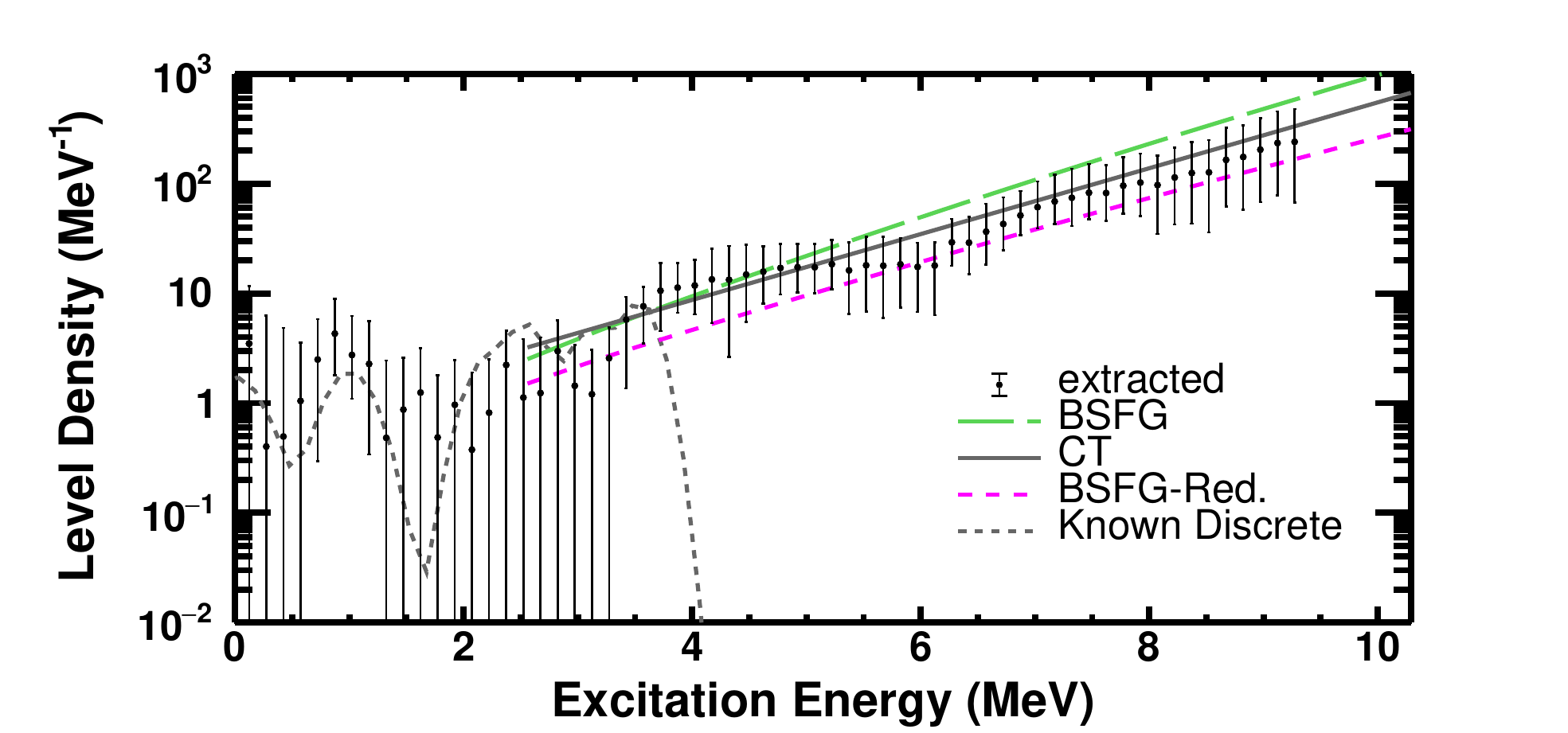}
    \caption{ The extracted $\rho_{\rm expt}$ for $^{60}{\rm Zn}$ (markers) compared to global theoretical $\rho$ models discussed in the text. The gray-dashed histogram is $\rho$ from known levels up to $E^{*}$ where all levels are thought to be known, where our experimental energy resolution has been applied.}
    \label{FinalLDvsGlobal}
\end{figure}

Our final $d\sigma(E)/dE$ for $^{58}{\rm Ni}(^{3}{\rm He},n)$ and the associated $\rho_{\rm expt}(E^{*})$ for $^{60}{\rm Zn}$ are shown in Figures~\ref{FinalCS} and \ref{FinalLDvsGlobal}, respectively. The contributions to the $\rho_{\rm expt}$ uncertainty, presented in Table~\ref{tab:LD}, include the statistical and systematic uncertainties from $d\sigma/dE|_{\rm expt}~\delta\rho_{\rm d\sigma/dE}$, a systematic uncertainty from the initial choice of $\rho_{\rm th}$ (estimated using $\rho_{\rm BSFG}$ and $\rho_{\rm BSFG-Red.}$ described below) $\delta\rho_{\rm model}$, a systematic uncertainty from the subtraction of non-statistical contributions to $d\sigma/dE$ (described in $\rho$-extraction step 2) $\delta\rho_{\chi^2}$, a systematic uncertainty from the normalization (described in $\rho$-extraction step 3) $\delta\rho_\mathcal{F}$, and a systematic uncertainty for adopting different $\sigma^{2}$ (described in $\rho$-extraction step 5) $\delta\rho_{\sigma^2}$. The dominant uncertainty contributions are statistical for low $E^{*}$ (high neutron energy) and the uncertainty from the initial choice of $\rho_{\rm th}$ for high $E^{*}$ (low neutron energy).

In principle, our extracted $\rho_{\rm expt}$ is also sensitive to the properties of the states in the discrete level region. The $J^{\pi}$ of several states are uncertain and some states have been observed by a single measurement, but not others~\cite{Brow13}. Rather than attempting to assign an associated uncertainty contribution, instead we present our $\rho_{\rm expt}$ and $\mathcal{F}$ in Table~\ref{tab:LD}. If future measurements modify the structure of $^{60}{\rm Zn}$ for $E^{*}<4$~MeV, calculations could be performed comparing the updated $d\sigma/dE|_{\rm th}$ to the result from this work in order to determine the required modification to $\mathcal{F}$. In principle, the correction to the measured $d\sigma/dE$ subtracting non-statistical components would also have to be repeated, but in practice this correction is not significant (see Figure~\ref{FinalCS}).

%%%%Discussion in paragraph below was for when erroneous x4 variation on sigma^2 was used.%%%%%
%We do not treat the impact of $\sigma^{2}$ as an additional uncertainty contribution, as our modification factor is somewhat speculative and we can remove this uncertainty with subsequent measurements, as discussed in Section~\ref{sec::level6}. Figures~\ref{fig:SpinImpactDiffCS} and \ref{fig:SpinImpactLD} show the impact of increasing or decreasing $\sigma^{2}$ by a factor of two or adopting the empirical form from Ref.~\cite{vonEgidy}. For an increase (only for $^{60}{\rm Zn}$), $\rho_{\rm expt}$ is enhanced above the discrete level region, while for $\sigma^{2}$ decrease or using the alternate parameterization, there is little impact. This can be understood by inspecting Figure~\ref{fig:SpinDist}, where is is shown that the increase in $\sigma^{2}$  brings $P(J)$ for the $E^{*}$ region populated in $^{58}{\rm Ni}(^{3}{\rm He},n)^{60}{\rm Zn}$ closer to $P(J)$ for the $E^{*}$ region populated in $^{61}{\rm Zn}$ by $^{58}{\rm Ni}+^{3}{\rm He}$. Therefore, enhancing $\sigma^{2}$ for $^{60}{\rm Zn}$ reduces the centrifugal barrier for neutron emission to higher $E^{*}$ relative to the barrier for emission to the discrete level region. Of course, if $\sigma^{2}$ were simultaneously enhanced for $^{61}{\rm Zn}$, then this relative barrier reduction would not happen and therefore $d\sigma/dE$, and subsequently $\rho_{\rm expt}$, would not be impacted.

\begin{table*}[!htbp]
\caption{\label{tab:LD}
Extracted $\rho_{\rm expt}$ for $^{60}{\rm Zn}$, total uncertainty $\delta\rho_{\rm tot}$, and uncertainty contributions in MeV$^{-1}$, where $\mathcal{F}=$0.795. Individual error bars are shown within the parentheses, the lower error bar being listed first, and the upper error bar following. If the lower error bar contains 0, the maximum level density is listed instead of the central value since all physical values below the upper limit are within the uncertainty.}
\def\arraystretch{0.95}
\begin{ruledtabular}
%\begin{tabular}{lcccccccc}
\begin{tabular}{cccccccc}
$E^{*} [\rm MeV]$ & $\rho_{\rm expt}$  & $\delta\rho_{\rm tot}$ & $\delta\rho_{\rm d\sigma/dE}$ & $\delta\rho_{\mathcal{F}}$ & $\delta\rho_{\chi{^2}}$ & $\delta\rho_{\sigma{^2}}$ & $\delta\rho_{\rm model}$\\ \hline% & DZ \\ \hline
0.125	&	$<$11.7	&	$(5.81, 8.19)$   &   $(5.74, 5.74)$   &   $(.655, .655)$   &   $(.099, .092)$   &	$(.036, .018)$   &   $(.000, 2.40)$\\
0.275	&	$<$6.30	&	$(4.97, 5.90)$   &   $(4.89, 4.89)$   &   $(.075, .075)$   &   $(.056, .052)$   &   $(.079, .040)$   &   $(.000, .972)$\\
0.425	&	$<$4.86	&	$(3.93, 4.36)$   &   $(3.92, 3.92)$   &   $(.093, .093)$   &   $(.020, .018)$   &   $(.015, .007)$   &   $(.000, .433)$\\
0.575	&	$<$3.57	&	$(2.07, 2.52)$   &   $(2.05, 2.05)$   &   $(.196, .196)$   &   $(.011, .010)$   &   $(.011, .020)$   &   $(.000, .441)$\\
0.725	&	2.50	&	$(2.20, 3.32)$   &   $(2.13, 2.13)$   &   $(.469, .469)$   &   $(.029, .027)$   &   $(.023, .044)$   &   $(.000, 1.10)$\\
0.875	&	4.30	&	$(2.51, 4.59)$   &   $(2.34, 2.34)$   &   $(.807, .807)$   &   $(.060, .056)$   &   $(.031, .059)$   &   $(.000, 2.06)$\\
1.025	&	2.75	&	$(1.65, 3.49)$   &   $(1.55, 1.55)$   &   $(.515, .515)$   &   $(.072, .066)$   &   $(.018, .009)$   &   $(.000, 1.85)$\\
1.175	&	2.28	&	$(1.94, 3.33)$   &   $(1.89, 1.89)$   &   $(.427, .427)$   &   $(.051, .048)$   &   $(.001, .000)$   &   $(.000, 1.40)$\\
1.325	&	$<$2.44	&	$(1.51, 1.96)$   &   $(1.49, 1.49)$   &   $(.091, .091)$   &   $(.022, .020)$   &   $(.018, .009)$   &   $(.000, .463)$\\
1.475	&	$<$2.58	&	$(1.40, 1.71)$   &   $(1.37, 1.37)$   &   $(.164, .164)$   &   $(.005, .005)$   &   $(.012, .024)$   &   $(.000, .301)$\\
1.625	&	$<$3.15	&	$(1.53, 1.90)$   &   $(1.49, 1.49)$   &   $(.234, .234)$   &   $(.001, .001)$   &   $(.023, .045)$   &   $(.000, .356)$\\
1.775	&	$<$1.79	&	$(1.20, 1.30)$   &   $(1.19, 1.19)$   &   $(.091, .091)$   &   $(.002, .002)$   &   $(.007, .014)$   &   $(.000, .093)$\\
1.925	&	$<$2.47	&	$(1.61, 1.50)$   &   $(1.48, 1.48)$   &   $(.181, .181)$   &   $(.016, .015)$   &   $(.005, .009)$   &   $(.109, .000)$\\
2.075	&	$<$1.90	&	$(1.93, 1.52)$   &   $(1.49, 1.49)$   &   $(.071, .071)$   &   $(.045, .042)$   &   $(.061, .031)$   &   $(.376, .000)$\\
2.225	&	$<$2.50	&	$(2.54, 1.68)$   &   $(1.63, 1.63)$   &   $(.154, .154)$   &   $(.067, .062)$   &   $(.081, .041)$   &   $(.818, .000)$\\
2.375	&	$<$4.58	&	$(4.59, 2.35)$   &   $(2.29, 2.29)$   &   $(.418, .418)$   &   $(.073, .068)$   &   $(.040, .020)$   &   $(2.23, .000)$\\
2.525	&	$<$3.84	&	$(3.88, 2.72)$   &   $(2.67, 2.67)$   &   $(.211, .211)$   &   $(.072, .067)$   &   $(.079, .040)$   &   $(1.12, .000)$\\
2.675	&	$<$3.95	&	$(3.98, 2.72)$   &   $(2.69, 2.69)$   &   $(.232, .232)$   &   $(.052, .048)$   &   $(.041, .020)$   &   $(1.23, .000)$\\
2.825	&	$<$5.68	&	$(4.08, 2.70)$   &   $(2.59, 2.59)$   &   $(.560, .560)$   &   $(.034, .032)$   &   $(.028, .053)$   &   $(1.40, .000)$\\
2.975	&	$<$3.38	&	$(3.40, 1.94)$   &   $(1.90, 1.90)$   &   $(.270, .270)$   &   $(.062, .057)$   &   $(.049, .025)$   &   $(1.44, .000)$\\
3.125	&	$<$3.07    &	$(3.12, 1.86)$   &   $(1.78, 1.78)$   &   $(.226, .226)$   &   $(.102, .095)$   &   $(.125, .063)$   &   $(1.21, .000)$\\
3.275	&	$<$4.89	&	$(3.95, 2.32)$   &   $(2.21, 2.21)$   &   $(.482, .482)$   &   $(.130, .121)$   &   $(.122, .061)$   &   $(1.57, .000)$\\
3.425	&	5.76	&	$(4.41, 3.51)$   &   $(3.24, 3.24)$   &   $(1.08, 1.08)$   &   $(.241, .223)$   &   $(.188, .094)$   &   $(.799, .000)$\\
3.575	&	7.62	&	$(4.14, 3.83)$   &   $(3.46, 3.46)$   &   $(1.43, 1.43)$   &   $(.269, .249)$   &   $(.118, .083)$   &   $(.271, .000)$\\
3.725	&	10.6	&	$(6.06, 8.38)$   &   $(4.11, 4.11)$   &   $(1.99, 1.99)$   &   $(.246, .228)$   &   $(.036, 2.92)$   &   $(1.45, .885)$\\
3.875	&	11.3	&	$(4.64, 7.53)$   &   $(4.08, 4.08)$   &   $(2.12, 2.12)$   &   $(.169, .156)$   &   $(.036, 1.47)$   &   $(.000, 1.46)$\\
4.025	&	11.8	&	$(5.36, 8.50)$   &   $(4.81, 4.81)$   &   $(2.21, 2.21)$   &   $(.154, .142)$   &   $(.066, 1.82)$   &   $(.000, 1.39)$\\
4.175	&	13.4	&	$(8.05, 12.2)$   &   $(7.56, 7.56)$   &   $(2.52, 2.52)$   &   $(.165, .153)$   &   $(.086, 2.85)$   &   $(.000, 1.42)$\\
4.325	&	13.3	&	$(10.6, 13.8)$   &   $(9.55, 9.55)$   &   $(2.50, 2.50)$   &   $(.172, .159)$   &   $(.089, 2.93)$   &   $(.686, .979)$\\
4.475	&	14.8	&	$(9.35, 12.9)$   &   $(8.14, 8.14)$   &   $(2.78, 2.78)$   &   $(.172, .159)$   &   $(.129, 3.25)$   &   $(.612, 1.07)$\\
4.625	&	15.8	&	$(7.69, 11.1)$   &   $(5.98, 5.98)$   &   $(2.96, 2.96)$   &   $(.173, .160)$   &   $(.151, 3.47)$   &   $(.870, .993)$\\
4.775	&	17.0	&	$(7.25, 11.1)$   &   $(5.35, 5.35)$   &   $(3.20, 3.20)$   &   $(.179, .166)$   &   $(.186, 3.77)$   &   $(.827, 1.08)$\\
4.925	&	17.3	&	$(7.11, 11.0)$   &   $(5.39, 5.39)$   &   $(3.25, 3.25)$   &   $(.217, .201)$   &   $(.191, 3.37)$   &   $(.629, 1.37)$\\
5.075	&	17.3	&	$(7.34, 10.8)$   &   $(5.65, 5.65)$   &   $(3.25, 3.25)$   &   $(.259, .240)$   &   $(.167, 2.62)$   &   $(.647, 1.65)$\\
5.225	&	18.5	&	$(7.55, 12.3)$   &   $(6.39, 6.39)$   &   $(3.47, 3.47)$   &   $(.302, .280)$   &   $(.169, 2.74)$   &   $(.099, 2.27)$\\
5.375	&	16.2	&	$(9.71, 13.0)$   &   $(7.53, 7.53)$   &   $(3.04, 3.04)$   &   $(.332, .307)$   &   $(.123, 3.26)$   &   $(1.46, 1.58)$\\
5.525	&	18.0	&	$(11.2, 14.9)$   &   $(9.38, 9.38)$   &   $(3.38, 3.38)$   &   $(.361, .334)$   &   $(.170, 2.91)$   &   $(1.02, 2.05)$\\
5.675	&	17.9	&	$(11.9, 14.8)$   &   $(9.48, 9.48)$   &   $(3.35, 3.35)$   &   $(.397, .367)$   &   $(.136, 2.76)$   &   $(1.68, 1.99)$\\
5.825	&	18.4	&	$(11.0, 13.6)$   &   $(8.09, 8.09)$   &   $(3.45, 3.45)$   &   $(.439, .406)$   &   $(.146, 2.74)$   &   $(2.01, 2.05)$\\
5.975	&	17.4	&	$(10.7, 11.6)$   &   $(6.73, 6.73)$   &   $(3.27, 3.27)$   &   $(.488, .452)$   &   $(.128, 2.42)$   &   $(3.05, 1.64)$\\
6.125	&	17.9	&	$(11.6, 11.3)$   &   $(6.95, 6.95)$   &   $(3.36, 3.36)$   &   $(.552, .511)$   &   $(.102, 1.81)$   &   $(3.74, 1.71)$\\
6.275	&	29.2	&	$(11.4, 18.7)$   &   $(9.42, 9.42)$   &   $(5.48, 5.48)$   &   $(.623, .577)$   &   $(.430, 2.69)$   &   $(.000, 5.05)$\\
6.425	&	29.0	&	$(14.0, 20.7)$   &   $(12.1, 12.1)$   &   $(5.44, 5.44)$   &   $(.734, .679)$   &   $(.388, 2.39)$   &   $(.387, 5.03)$\\
6.575	&	36.5	&	$(18.2, 28.9)$   &   $(16.2, 16.2)$   &   $(6.85, 6.85)$   &   $(.888, .822)$   &   $(.592, 3.90)$   &   $(.000, 7.41)$\\
6.725	&	42.9	&	$(18.2, 31.8)$   &   $(15.5, 15.5)$   &   $(8.05, 8.05)$   &   $(1.09, 1.01)$   &   $(.712, 4.76)$   &   $(.000, 9.58)$\\
6.875	&	52.3	&	$(17.3, 34.3)$   &   $(13.3, 13.3)$   &   $(9.63, 9.63)$   &   $(1.34, 1.24)$   &   $(.872, 5.42)$   &   $(.000, 12.5)$\\
7.025	&	61.1	&	$(22.1, 43.8)$   &   $(17.5, 17.5)$   &   $(11.5, 11.5)$   &   $(1.67, 1.55)$   &   $(1.11, 6.93)$   &   $(.000, 15.9)$\\
7.175	&	68.8	&	$(26.0, 52.2)$   &   $(21.0, 21.0)$   &   $(12.9, 12.9)$   &   $(2.09, 1.94)$   &   $(1.21, 8.48)$   &   $(.000, 18.9)$\\
7.325	&	74.3	&	$(33.3, 62.3)$   &   $(28.9, 28.9)$   &   $(14.0, 14.0)$   &   $(2.60, 2.41)$   &   $(1.15, 8.67)$   &   $(.000, 21.4)$\\
7.475	&	82.6	&	$(35.4, 68.2)$   &   $(30.4, 30.4)$   &   $(15.5, 15.5)$   &   $(3.21, 2.98)$   &   $(1.11, 8.79)$   &   $(.000, 25.1)$\\
7.625	&	82.1	&	$(36.5, 65.5)$   &   $(27.8, 27.8)$   &   $(15.4, 15.4)$   &   $(3.90, 3.61)$   &   $(.583, 7.08)$   &   $(3.88, 26.4)$\\
7.775	&	95.7	&	$(42.8, 77.7)$   &   $(31.3, 31.3)$   &   $(18.0, 18.0)$   &   $(4.69, 4.34)$   &   $(.554, 8.75)$   &   $(5.85, 32.6)$\\
7.925	&	102.	&	$(51.5, 86.0)$   &   $(34.3, 34.3)$   &   $(19.2, 19.2)$   &   $(5.47, 5.06)$   &   $(.123, 9.49)$   &   $(11.7, 36.8)$\\
8.075	&	97.1	&	$(62.4, 83.8)$   &   $(35.1, 35.1)$   &   $(18.2, 18.2)$   &   $(6.27, 5.80)$   &   $(.000, 6.37)$   &   $(22.4, 37.4)$\\
8.225	&	114.    &	$(71.3, 98.4)$   &   $(39.8, 39.8)$   &   $(21.4, 21.4)$   &   $(7.02, 6.50)$   &   $(.000, 6.63)$   &   $(25.6, 46.1)$\\
8.375	&	125. 	&	$(81.7, 113.)$   &   $(44.1, 44.1)$   &   $(23.5, 23.5)$   &   $(7.62, 7.06)$   &   $(.000, 9.33)$   &   $(31.2, 53.6)$\\
8.525	&	127. 	&	$(90.6, 122.)$   &   $(45.5, 45.5)$   &   $(23.8, 23.8)$   &   $(8.11, 7.51)$   &   $(.000, 12.5)$   &   $(38.6, 57.8)$\\
8.675	&	165. 	&	$(103., 159.)$   &   $(52.2, 52.2)$   &   $(30.9, 30.9)$   &   $(8.42, 7.79)$   &   $(.000, 20.3)$   &   $(41.5, 77.8)$\\
8.825	&	175. 	&	$(117., 164.)$   &   $(56.1, 56.1)$   &   $(32.8, 32.8)$   &   $(8.78, 8.13)$   &   $(.491, 10.7)$   &   $(50.9, 87.9)$\\
8.975	&	204. 	&	$(136., 190.)$   &   $(63.0, 63.0)$   &   $(38.3, 38.3)$   &   $(8.96, 8.30)$   &   $(1.99, 7.74)$   &   $(59.9, 108.)$\\
9.125	&	234. 	&	$(156., 222.)$   &   $(69.2, 69.2)$   &   $(44.0, 44.0)$   &   $(8.90, 8.24)$   &   $(3.96, 10.2)$   &   $(69.9, 129.)$\\
9.275	&	240. 	&	$(173., 239.)$   &   $(73.1, 73.1)$   &   $(45.1, 45.1)$   &   $(8.66, 8.02)$   &   $(9.13, 12.7)$   &   $(77.8, 140.)$\\
\end{tabular}
\end{ruledtabular}
\end{table*}

In Figure~\ref{FinalLDvsGlobal} we compare $\rho_{\rm expt}$ to arguably the two most common phenomenological $\rho$: the CT model~\cite{Gilb65} (Eqn.~\ref{eqn:CT}~\footnote{Note that in our {\tt TalysV1.8} calculations we are actually employing the constant-temperature + Fermi-gas model; however, the transition to the Fermi gas portion of $\rho$ does not happen for $^{60}{\rm Zn}$ until $E^{*}=16.9$~MeV and is therefore not relevant for this work.}) and the back-shifted Fermi gas (BSFG) model~\cite{Dilg}. For the latter, 
\begin{equation}
\rho_{\rm{BSFG}}(E^{*})=\frac{\exp{\left(2\sqrt{a(E^{*}-\Delta)}\right)}}{12\sqrt{2}\sigma a^{1/4}(E^{*}-\Delta)^{5/4}},
\label{eqn:rhobsfg}
\end{equation}
where $\Delta$ is a back-shift to ensure $\rho$ is described at low-lying excitation energies where all discrete levels are known and $a$ is the $E^{*}$-dependent level-density parameter of Ref.~\cite{Igna75}, based on global fits to $\rho$. For both the the CT and BSFG models, we use the parameterization and default parameters of {\tt TalysV1.8}. Based on an analysis of $\rho$ determined from low-lying excited states, Refs.~\cite{AlQur01,AlQur} found that $a$ may be reduced for nuclides away from the valley of $\beta$-stability. A favored reduction was the form
\begin{equation}
 a_{\rm red}=\frac{\alpha{A}}{\exp\left(\gamma\left(Z-Z_0\right)^2\right)},
    \label{eqn: bsfgred}
\end{equation}
where 
%$\alpha$=0.1068, $\gamma$=0.0389, and $Z_0$=0.5042$A/(1+0.00732A^{2/3})$. 
$\alpha$, $\gamma$, and $Z_{0}$ were determined empirically and, for $^{60}{\rm Zn}$, the result is $a_{\rm red}=4.73$. We therefore explore adjusting $a(E^{*})$ such that $a(S_{n})=a_{\rm red}$, referring to this calculation as BSFG-Red.

\begin{figure}
    \centering
    \includegraphics[width=\columnwidth]{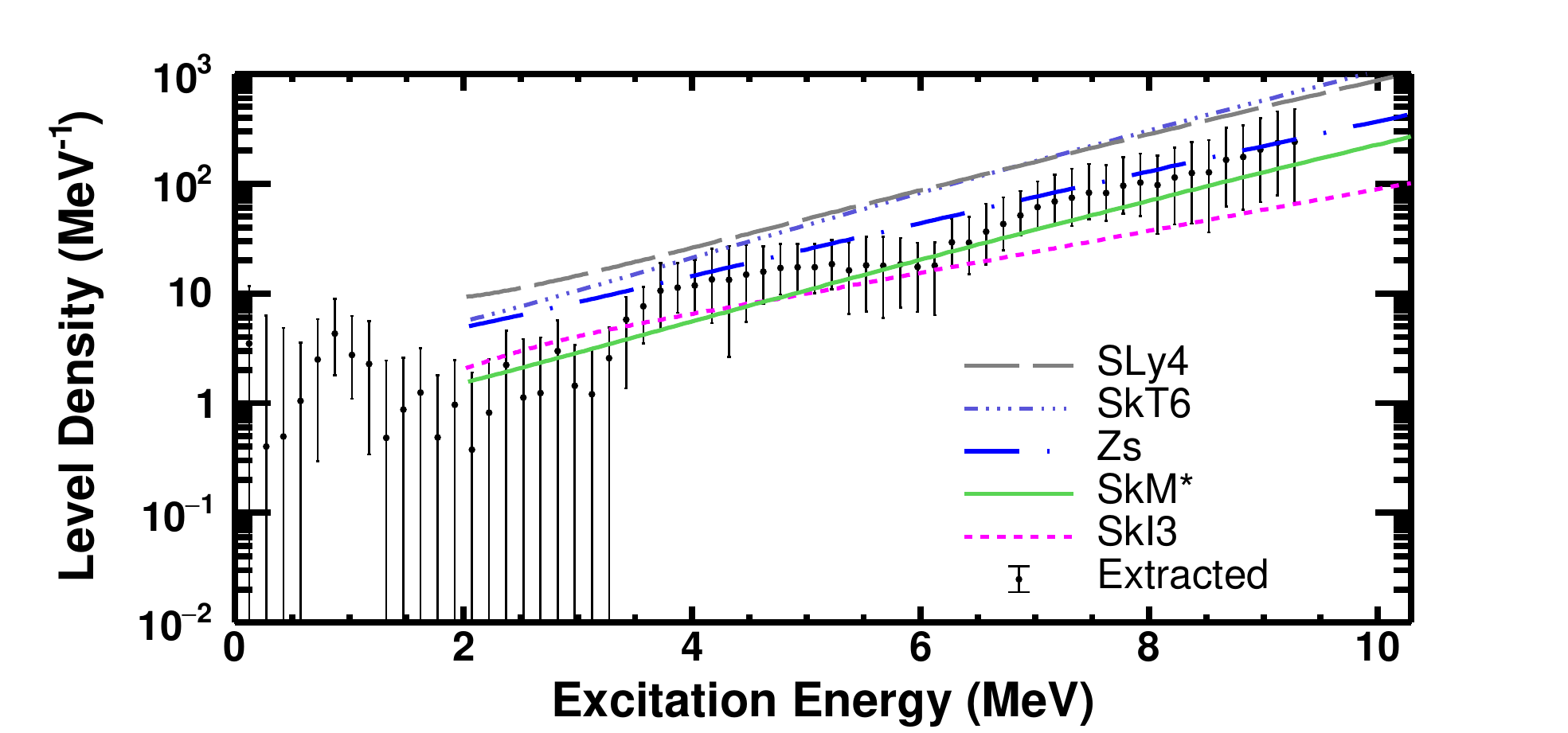}
    \caption{Extracted $^{60}{\rm Zn}$ $\rho$ (markers) compared to microscopic theoretical $\rho$ models discussed in the text.}
    \label{FinalLDvsMicro}
\end{figure}

In Figure~\ref{FinalLDvsMicro} we compare our results to microscopic level densities. These were obtained by generating single particle energy levels using various energy density functionals within the Hartree Fock Bogoliubov framework~\cite{Mamu18} and then determining $\rho$ using statistical mechanics~\cite{Sano63,Mamu15}. We adopted energy density functionals based on the SLy4~\cite{Chab98}, SkT6~\cite{Tond84}, Zs~\cite{Frie86}, SkM$^{*}$~\cite{Bart82}, and SkI3~\cite{Rein95} Skyrme forces as a somewhat arbitrary sampling of the vast array of possible choices. Note that we do not report microscopic $\rho$ below 2~MeV as the saddle-point approximation used in the $\rho$ calculation diverges. Such divergence can be corrected with asymptotic expressions, but this was not deemed necessary here.

\begin{figure}
    \centering
    \includegraphics[width=\columnwidth]{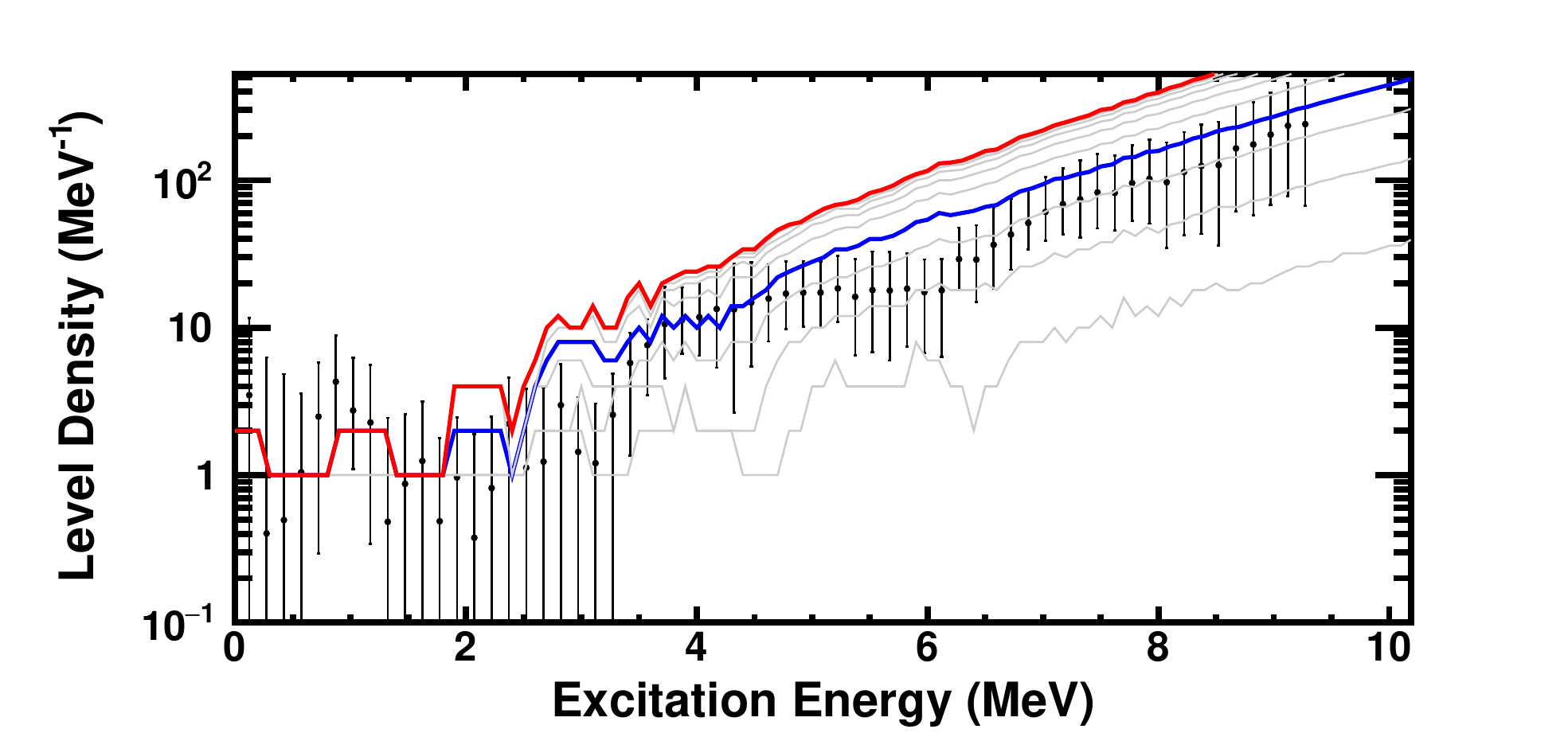}
    \caption{Extracted $^{60}{\rm Zn}$ $\rho$ (markers) compared to the shell model based theoretical $\rho$ model discussed in the text (ELM). Lines are $\rho$ summed up to a maximum $J$ from $J=0$ (lowest line) up through $J=12$ (red, upper bold line), where the sum up through $J=3$ is the lower bold line, indicated in blue.}
    \label{FinalLDvsShell}
\end{figure}

In Figure~\ref{FinalLDvsShell} we compare our results to calculations performed with the extrapolated Lanczos method (ELM) of Ref.~\cite{Orma19}. This is a shell-model based $\rho$, using the pf-shell model space and the GXPF1A Hamiltonian~\cite{Honm05}. The ELM $\rho$ are summed for all positive-parity states up through a maximum $J$. For $J$=0$-$8, we use 8 moments and 30 iterations for each $J$, known as ELM(8,30), whereas ELM(8,4) is used for higher $J$.

\section{\label{sec::level5} Discussion}

\subsection{Comparison to Level Density Models}
\label{ssec:LDcomparison}
We compare our results to phenomenological, microscopic, and shell-model based theory calculations of $\rho$, as each technique has its advantages. Each comparison also provides unique context for understanding $\rho$ of $^{60}{\rm Zn}$ and likely nearby nuclides, which is valuable for predictions of statistical properties of nuclides off-stability for astrophysics and applications.

Our comparison to phenomenological $\rho$ in Figure~\ref{FinalLDvsGlobal} demonstrates that we clearly favor the CT model of {\tt TalysV1.8}. This adds to the body of evidence that the CT $\rho$ is often favored over the BSFG prediction for nuclides in this mass region~\cite{Voin09,Ramirez13} (and perhaps all mass regions~\cite{Gutt15}). The agreement may be explained by the conclusion that the CT model effectively captures the phase transition from ordered nuclear structure to quantum chaos with increasing $E^{*}$~\cite{Zele18}. While a reduced $a$ within the BSFG model may improve agreement with our results, we find that the required $a$-reduction is not consistent with $a_{\rm red}$ predicted by Refs.~\cite{AlQur01,AlQur}.

The comparison to microscopic $\rho$ in Figure~\ref{FinalLDvsMicro} demonstrates that these theoretical $\rho$ strongly depend on the adopted Skyrme force. In this case, Zs is clearly favored. However, we note that normalization has not been performed to match $\rho$ in the discrete level region, as is often done for databases of $\rho$ intended for HF calculations. Nonetheless, this may provide guidance for microscopic calculations of statistical nuclear properties for nuclides in this region of the nuclear chart.

Figure~\ref{FinalLDvsShell} shows that the shell-model based ELM calculations significantly overestimate $\rho$ for $^{60}{\rm Zn}$, while Figure~\ref{fig:SpinDist} shows general agreement between the ELM $P(J)$ and commonly used theoretical estimates. The ELM method has previously been demonstrated to successfully reproduce measured $\rho$; however, results are somewhat dependent on the adopted Hamiltonian~\cite{Orma19}. While shell model calculations face the usual limits imposed by a finite model space, this would only artificially reduce $\rho$ at high $E^{*}$. Alternatively, the disagreement between our result and the ELM results could be explained if our measurement failed to probe $J>3$. However, this would require significantly reduced $\mathcal{T}_{n}$ for large angular momentum $\ell$ transfer neutron emissions. 
%relative to the barrier associated with the $\mathcal{T}_{n}$ used here.

It is interesting to note that our $\rho_{\rm expt}$ exhibits a plateau from $E^{*}\sim5-6$~MeV. This is counter to the usual expectation of exponential growth in $\rho$ with increasing $E^{*}$ and is not seen in any of the $\rho$ models that we compare to. At present, we do not have an explanation for this feature, though we note something similar is seen at $E^{*}\sim3-4$~MeV for $^{57}{\rm Fe}$~\cite{VoinovGe}. We note that though exponential $\rho_{\rm expt}$ growth for $E^{*}\sim5-6$~MeV appears to be within uncertainties, the uncertainty in this region is dominated by correlated systematics. Here, the uncertainty of the shape of the $\rho_{\rm expt}$ trend is primarily due to statistical uncertainties, which are much smaller than the total error bars.

\subsection{Implications for Astrophysics}

The importance of $^{60}{\rm Zn}$ to the $rp$-process and Type-I X-ray bursts is discussed in Section~\ref{sec:level1}. In calculating $^{59}{\rm Cu}(p,\gamma)^{60}{\rm Zn}$ via the HF-formalism, it should be noted that the uncertainties in the $^{59}{\rm Cu}+p$ proton optical potential (at low temperature) and the $^{60}{\rm Zn}$ $\gamma$-strength function (at high temperature) are far more significant contributors to the astrophysical reaction rate uncertainty than the $^{60}{\rm Zn}$ $\rho$~\citep{Raus12}.
%, when considering the available $\gamma$-strength function and $\rho$ models available in {\tt TalysV1.8}. 
However, $\rho$ can help answer the key question as to whether the HF approach is likely a valid approximation for this reaction. In particular, we are interested in whether the $E^{*}$ region populated in the X-ray burst environment has a sufficiently high $\rho$.

\begin{figure}
    \centering
    \includegraphics[width=\columnwidth]{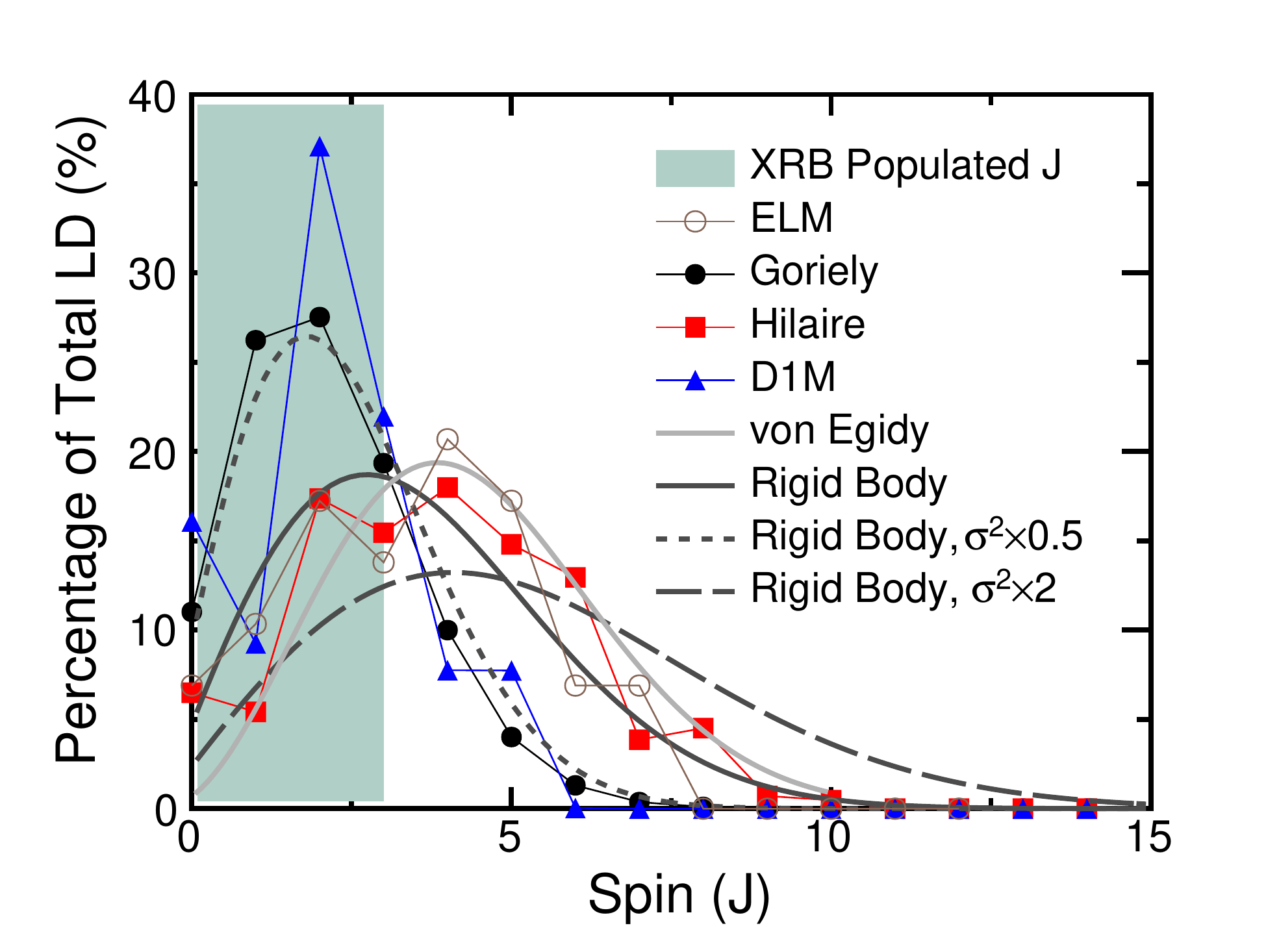}
    \caption{Calculated $P(J)$ of $^{60}{\rm Zn}$ at $E^{*}=5$~MeV using the nominal $\sigma_{\rm FG}^{2}$ (``Rigid Body"), $\sigma_{\rm FG}^{2}$ increased or decreased by a factor of two, or $\sigma^{2}$ from Ref.~\cite{vonEgidy} (``von Egidy"), compared to ELM $P(J)$ and microscopic $P(J)$ available in {\tt TalysV1.8}. The green band indicates the relevant $J$ for $^{59}{\rm Cu}(p,\gamma)^{60}{\rm Zn}$ in a Type-I X-ray burst (XRB) environment.}
    \label{fig:SpinDist}
\end{figure}

For the relevant temperature range, $0.5-1.0$~GK, $^{59}{\rm Cu}(p,\gamma)$ will populate $E^{*}\approx5.6-6.3$~MeV in $^{60}{\rm Zn}$~\cite{Raus10}. This happens to correspond to the near-plateau present in our data in Figure~\ref{FinalLDvsGlobal} and Table~\ref{tab:LD}, where $\rho$ maintains $\sim20$~MeV$^{-1}$, contrary to the exponential growth present in theory calculation results. While this is above the $\rho\sim10$~levels-per-astrophysical-window (here $\sim$12 window$^{-1}$) heuristic for the applicability of HF results~\cite{Rauscher97}, it is low enough that in practice HF-applicability will depend on $\sigma^{2}$. This is because, in the X-ray burst environment, the only relevant states in $^{60}{\rm Zn}$ are those populated by proton-captures involving low $\ell$ transfer. Assuming only $\ell=0$ and $\ell=1$ proton-captures are relevant, then roughly half of the total $\rho$ is encompassed for the nominal choice of $\sigma^{2}$ (See Figure~\ref{fig:SpinDist}). However, if $\sigma^{2}$ is doubled, following the systematics of Ref.~\cite{Grim16}, only around a third of all levels are within the relevant $J$ range. Therefore, a measurement of $\sigma^{2}$ is needed to confirm the applicability of HF calculations for the $^{59}{\rm Cu}(p,\gamma)^{60}{\rm Zn}$ reaction rate in X-ray burst conditions. In this regard, the parity ratio~\citep{AlQur} may also be of interest, though we do not examine its impact here.

\section{\label{sec::level6} Conclusion}

Using neutron evaporation spectra from $^{58}{\rm Ni}(^{3}{\rm He},n)$, we determined $\rho$ for $^{60}{\rm Zn}$ in the $E^{*}$ range of astrophysical interest and compared our results to several theoretical predictions, including phenomenological, microscopic, and shell-model based calculations. We find that the CT model of {\tt TalysV1.8} is in best agreement with our data amongst phenomenological models and that there is no evidence for a $\rho$ reduction moving away from the valley of $\beta$-stability. Of our microscopic $\rho$ calculations, the results from using the Zs Skyrme force most closely resemble our experimental results.  Our ELM $\rho$ results are substantially larger than our experimental results, indicating either a limitation of the shell model approach here, or that, contrary to expectations, our experiment failed to probe high-$J$ levels. For $E^{*}$ relevant for the $^{59}{\rm Cu}(p,\gamma)^{60}{\rm Zn}$ reaction rate at X-ray burst temperatures, we find a plateau in $\rho$, where the exponential growth with increasing $E^{*}$ is temporarily halted. Depending on the adopted $\sigma^{2}$, it is possible that the HF formalism may not be applicable for this reaction rate. To resolve this issue, as well as the discrepancy with ELM results, we suggest measurements of $\sigma^{2}$ and perhaps high-resolution spectroscopy for $E^{*}\sim6$~MeV in $^{60}{\rm Zn}$.

\begin{acknowledgements}
 This work was supported in part by the U.S. Department of Energy under Grant No. DE-FG02-88ER40387, DE-NA0003883, DE-NA0003909, and DE-SC0019042. We benefited from support by the National Science Foundation under Grant No. PHY-1430152 (JINA Center for the Evolution of the Elements). Part of this work was performed under the auspices of the U.S. Department of Energy by Lawrence Livermore National Laboratory under Contract DE-AC52-07NA27344 and DOE/SC/NP Field Work Proposal SCW0498.
\end{acknowledgements}

\bibliographystyle{apsrev4-2}
\bibliography{CuBib}% Produces the bibliography via BibTeX.
\end{document}